# Topological Electronic States and Phonon-Mediated Superconductivity in Ru-Based Ternary Pnictides: ZrRuAs and HfRuP

**Jubair Hossan Abir[1], Tanvir Khan[1], Tauhidur Rahman[1], Md. Kamrul Hassan[1], Sraboni Saha Moly[1], Mst. Maskura Khatun[1], Raihana Shams Islam[1*], Saleh Hasan Naqib[1*]**

[1]Department of Physics, University of Rajshahi, Rajshahi 6205
*Corresponding authors; Emails: rsislam@ru.ac.bd, salehnaqib@yahoo.com

## Abstract

Topological quantum materials have emerged as a promising candidate for next-generation technologies due to their protected surface states and exotic quasiparticles excitations. Among them, topological superconductors can enable quantum computing through intrinsically protected Majorana-based qubits. This study investigates the structural, electronic, mechanical, optical, thermodynamic, and superconducting properties of ZrRuAs and HfRuP. Electronic band structure calculations reveal topological semimetallic behavior in both compounds with several bands crossing the Fermi level. The inclusion of spin-orbit coupling lifts band degeneracies in both materials. The Fermi surface analysis shows both electron and hole like sheets. The estimated Pugh's ratio, Poisson's ratio and machinability index indicate that both compounds are ductile in nature and exhibits excellent machinability. Optical results show strong metallic reflectivity and notable absorption in the ultraviolet region. Thermodynamic analysis indicates stable behavior. The estimated superconducting transition temperatures are 9.75 K for ZrRuAs and 9.03 K for HfRuP that indicates medium-coupling conventional superconductivity. The results provide a detailed view of the physical and superconducting properties of ZrRuAs and HfRuP and help to close the existing research gap in these topological superconducting materials.

**Keywords:** Topological quantum materials; Topological superconductor; Ru-based ternary pnictides; Thermophysical properties; Optoelectronic properties

## 1 Introduction

Topological superconductors (TSCs) have become a major focus of condensed-matter research because of their exceptional electronic states and their relevance to quantum phenomena [1–6]. Despite sustained progress in both theory and experiment, the number of materials identified as possible bulk TSCs remains quite limited. To date, only a few systems, such as $Sr_2RuO_4$ [7], the Weyl semimetal Td-$MoTe_2$ [8], and $PdTe_2$ [9] have been widely discussed as promising candidates. In many-body quantum systems, states that can be smoothly reduced to a simple atomic-limit description are classified as topologically trivial. By contrast, if a system cannot be adiabatically connected to this atomic limit, it is regarded as topologically nontrivial. The quantum Hall effect was the first physical example identified as a topological quantum state. In that system, the relevant topology is described by an integer known as the Chern number [10,11]. Although the topological nature of the quantum Hall effect had already been identified in 1982 [11], the field of topological quantum matter attracted much broader attention after 2005. In that year, theory predicted the quantum spin Hall insulator phase in a two-dimensional time-reversal-symmetric insulating system [12,13]. The same concept was later generalized to three-dimensional materials [14–16], which led to the introduction of the term topological insulator [14]. A brief historical overview of topological insulators can be found in a recent review article [17]. It was later understood that topological classification could, in principle, be applied to many kinds of gapped quantum many-body systems, where the energy gap protects the occupied states. Consequently, researchers developed systematic classifications not only for insulating phases but also for superconducting phases [17,18]. These classification schemes showed that the possible topological invariants of a gapped system are determined by its symmetry class and spatial dimensionality. For instance, a superconductor that breaks time-reversal symmetry can exhibit nontrivial topology in two-dimensions (2D), but not in three-dimensions (3D).

In the search for topological superconductors, ternary transition-metal pnictides with the general formula TT′X have attracted considerable attention [19–23]. In these family of compounds, T may be Ca, Zr, or Hf; T′ may be Ir, Ru, Ag, or Os; and X may be P, As, or Si. These materials are particularly interesting because several members show nontrivial topological electronic properties and some of them also exhibit superconductivity. The Ru-based compounds received increased attention after relatively high superconducting transition temperatures were reported for ZrRuP ($T_c$ = 13.0 K) and HfRuP ($T_c$ = 12.7 K) [22,23]. This discovery encouraged further studies on related Zr- and Hf-based compounds. So far, research on these materials has mainly focused on their superconducting properties. Although oxide superconductors attracted enormous interest after the discovery of high-temperature superconductivity [21], intermetallic superconductors remain important from both scientific and technological perspectives [19,20,24]. This is because oxide superconductors often have brittle crystal structures [25,26], making them difficult to process into practical forms. In comparison, intermetallic superconductors are generally more suitable for fabrication and are widely relevant for applications such as superconducting wires, coils, and high-field magnets [27–34]. Among the intermetallic TT′X superconductors, ZrRuAs is especially notable because it shows a relatively high superconducting transition temperature of around 12 K along with a large upper critical field, $H_{c_2}$ = 14.1 T [21,35,36]. A comparative study by Shirotani and co-workers examined ZrRuAs alongside the isostructural compounds ZrRuP, HfRuP, and ZrRuSi [37–40]. Their results showed that all of these hexagonal ZrNiAl-type materials superconduct above 10 K, despite noticeable differences in unit-cell volume. This observation indicates that superconductivity in this structural family is not strongly controlled by lattice volume. They also found that ZrRuSi, although possessing one fewer valence electron than HfRuP, ZrRuP, and ZrRuAs, exhibits a transition temperature that remains close to those of the other compounds. Such behavior suggests that superconductivity in these hexagonal ZrNiAl-type systems is only weakly affected by valence-electron count, even though electron count often plays a decisive role in the transport and electronic properties of many other materials.

In the present study, we investigated the normal-state, superconducting-state, and topological properties of ZrRuAs and HfRuP compounds. The schematic structure is shown in **Figure 1.** Although numerous studies have previously explored its superconducting and topological characteristics, several important physical properties such as elastic behavior, bonding nature, thermodynamic properties, and optical response have not yet been studied in detail, to the best of our knowledge. We also re-examined its superconducting and topological features, and our findings are consistent with previous experimental reports. The main objective of this work is therefore to address the remaining research gaps associated with this material.

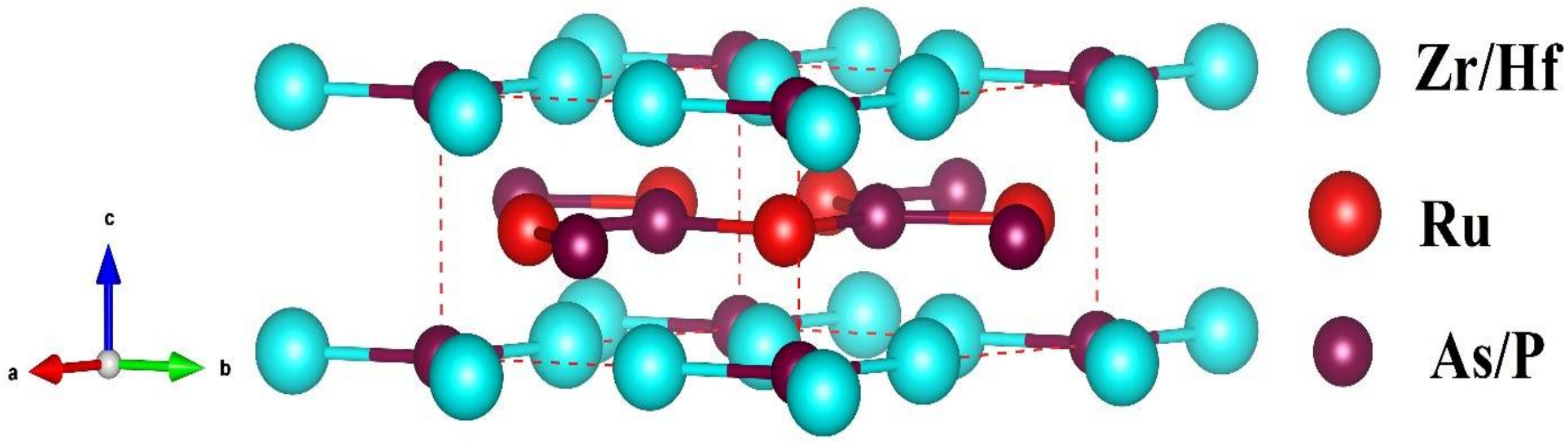

**Figure 1.** Schematic 3D representation of layered Crystal structures of ZrRuAs and HfRuP.

## 2 Computational Methodology

Density functional theory (DFT) computations were carried out using the Quantum ESPRESSO package [41,42]. The exchange-correlation contribution was treated using the Perdew-Burke-Ernzerhof form of the generalized gradient approximation (PBE-GGA) [43]. The Kohn-Sham wave functions were represented with a plane-wave basis, and the cutoff energy was fixed at 60 Ry. Interactions between the valence electrons and ionic cores were described using norm-conserving pseudopotentials [44]. To calculate the charge density convergence, the Methfessel-paxton smearing method was used with a smearing parameter of $\sigma = 0.01$ Ry. The Brillouin zone was sampled using Monkhorst-Pack *k*-point meshes of $12 \times 12 \times 12$ for structural optimization and $24 \times 24 \times 24$ for electronic-structure calculations. Geometry optimization was continued until the forces on atoms were reduced below $10^{-5}$ Ry/Å.

The mechanical behavior, electronic bonding nature, and optical properties of ZrRuAs and HfRuP were further examined using the CASTEP code [45]. In this work, the Brillouin zone (BZ) was sampled using a $10 \times 10 \times 15$ *k*-point grid, and the plane-wave kinetic energy cutoff was chosen as 400 eV after taking several runs with different energy cutoffs. The elastic constants were calculated by applying the stress-strain method [46]. The optical behavior of the compounds was studied over a range of photon energies from the complex dielectric function, written as $\varepsilon(\omega) = \varepsilon_1(\omega) + i\varepsilon_2(\omega)$. Once the dielectric function had been evaluated, the remaining optical parameters, such as the refractive index $n(\omega)$, absorption coefficient $\alpha(\omega)$, energy-loss function $L(\omega)$, reflectivity $R(\omega)$, and optical conductivity $\sigma(\omega)$, were calculated using the standard optical relations [47,48].

The influence of temperature and pressure on the thermodynamic, elastic, and structural behavior was examined by fitting the calculated energy-volume (*E*-*V*) data to the third-order Birch-Murnaghan equation of state (*E*-*V*) [49]. Thermodynamic parameters were evaluated using the GIBBS2 program [50] over a temperature range from 0 to 1000 K and a pressure range from 0 to 50 GPa.

Phonon dispersion relations were calculated using density functional perturbation theory (DFPT) [51]. For the electron-phonon coupling (EPC) calculations, a denser 24 × 24 × 24 *k*-point grid was employed, together with 4 × 4 × 4 and 3 × 3 × 3 *q*-point grids for ZrRuAs and HfRuP, respectively. The phonon calculations were carried out without including spin-orbit coupling, since its effect on the vibrational behavior and superconducting properties of these compounds is considered insignificant [52,53]. The surface-state spectra were obtained through the iterative Green's function approach available in the WANNIERTOOLS package [54,55]. For this purpose, the tight-binding Hamiltonian was generated from maximally localized Wannier functions [56,57].

# 3 Results and Discussion

## 3.1 Structural Properties and Electron Density Difference

The ternary transition-metal pnictides ZrRuAs and HfRuP crystallize in the hexagonal $P\bar{6}2m$(No. **189**) structure. In ZrRuAs, Zr and Ru atoms occupy the 3*f* (0.58333, 0, 0) and 3*g* (0.24478, 0, 0.5) Wyckoff positions, respectively, while As1 and As2 are located at the 2*d* (0.333, 0.667, 0.5) and 1*a* (0, 0, 0) sites. For HfRuP, Hf atoms occupy the 3*f* (0.58531, 0, 0), Ru atoms the 3*g* (0.24710, 0, 0.5) whereas P1 and P2 occupy the 1*a* (0, 0, 0) and 2*d* (0.333, 0.667, 0.5) positions, respectively.

**Table 1.** The optimized lattice parameters *a*(Å) and *c*(Å), and optimized cell volume *V*(Å$^3$) of the compounds.

| Compound | Functionals | *a* | *c* | *V* | Ref. |
|---|---|---|---|---|---|
| ZrRuAs | ----- | 6.586 | 3.888 | 145.863 | [58] |
| | GGA-PBESOL | 6.576 | 3.885 | 145.473 | This work |
| HfRuP | ----- | 6.414 | 3.753 | 133.711 | [58] |
| | GGA-PBESOL | 6.394 | 3.758 | 133.037 | This work |

**Figure 2. (a) and (b)** illustrate the 2D electron density difference (EDD) contour maps of the Ru-Zr-As and Ru-Hf-P systems, respectively. The color scale represents the degree of electron accumulation and depletion within the crystal structure. In the contour maps, the blue regions correspond to electron accumulation, whereas the yellow/red regions indicate electron depletion, demonstrating the redistribution of charge caused by chemical bonding and orbital hybridization among the constituent atoms.

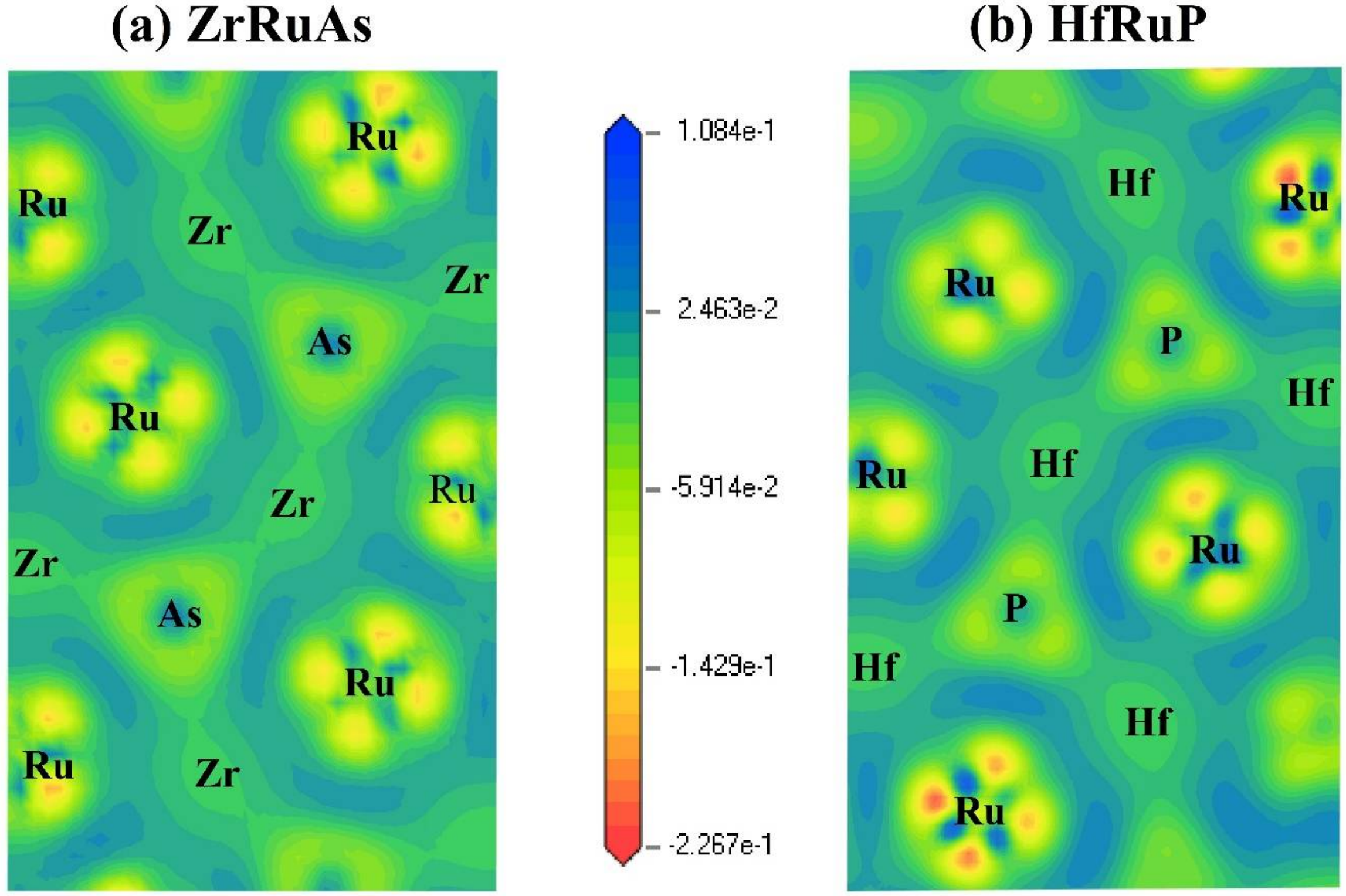

**Figure 2.** The electron density difference maps for (a) ZrRuAs and (b) HfRuP.

In **Figure 2(a),** significant charge redistribution is observed around the Ru atoms, indicating strong hybridization between the Ru *d* orbitals and the neighboring Zr *d* and As *p* orbitals. The electron accumulation around the Ru atoms indicates that the Ru-As bonds have covalent nature. On the other hand, the charge distribution around the Zr atoms is more spread out which suggests a delocalized type of interaction. The asymmetric distribution of charge density between Ru and As atoms shows that electrons are transferred from Zr to the Ru-As framework. This electron transfer helps to stabilize the crystal structure and also improves the electronic conductivity of the material. The continuous charge-density contours between neighboring atoms indicate noticeable orbital overlap and strong interactions among the atoms.

In **Figure 2(b)**, the electron density distribution shows a similar but comparatively stronger localization around the Ru atoms due to the presence of Hf and P atoms. The Hf atoms exhibit moderate electron depletion which indicates that Hf acts primarily as an electron donor within the lattice. In contrast, the P atoms show localized charge accumulation resulting from their higher electronegativity. The enhanced electron concentration around Ru and P suggests strong Ru-P covalent hybridization. It originates from the interaction between Ru 4*d* and P 3*p* states. Compared with the ZrRuAs, the HfRuP exhibits slightly stronger charge localization owing to the heavier atomic number and larger spin-orbit interaction associated with Hf atoms.

The calculated electron density difference maps suggest that both systems possess mixed metallic-covalent bonding characteristics. The metallic nature arises from the delocalized electron cloud distributed throughout the lattice while the directional charge accumulation between specific atomic pairs confirms covalent bonding contributions. Such bonding behavior is beneficial for structural stability, charge transport, and electronic performance. The observed charge transfer and orbital hybridization also indicate strong chemical interactions among the constituent atoms. It may play a crucial role in determining the electronic, catalytic and electrochemical properties of these materials.

## 3.2 Mechanical and Elastic Properties

Elastic constants help to examine important mechanical features such as structural stability, rigidity, deformation resistance, bonding directionality, and ductile behavior. Therefore, elastic analysis plays a key role in judging whether a crystal structure is mechanically stable and how it responds to external forces. For the hexagonal ZrRuAs and HfRuP compounds, the elastic response is fully characterized by five independent constants: $C_{11}$, $C_{12}$, $C_{13}$, $C_{33}$, and $C_{44}$. The sixth elastic constant, $C_{66}$, is not independent and can be obtained using,

$$C_{66} = \frac{(C_{11} - C_{12})}{2} \quad (1)$$

The calculated elastic constants are presented in **Table 2.** The values obtained satisfy all the mechanical stability criteria [59], confirming that both ZrRuAs and HfRuP are mechanically stable.

**Table 2.** Calculated elastic constants, $C_{ij}$ (GPa), tetragonal shear modulus, $C_t$ (GPa) and Cauchy pressure, $C_P$ (GPa).

| Compounds | $C_{11} = C_{22}$ | $C_{33}$ | $C_{44} = C_{55}$ | $C_{66}$ | $C_{12}$ | $C_{13}$ | $C_t$ | $C_P$ | Ref. |
|---|---|---|---|---|---|---|---|---|---|
| ZrRuAs | 317.47 | 327.71 | 99.62 | 98.86 | 119.75 | 151.02 | 98.86 | 20.13 | This work |
| HfRuP | 348.40 | 389.33 | 114.78 | 106.87 | 134.66 | 203.38 | 106.87 | 19.88 | This work |

The elastic constants $C_{11}$, $C_{22}$, and $C_{33}$ measure the crystal's resistance to linear strain along the crystallographic *a*-, *b*-, and *c*-axes, respectively. For these compounds, $C_{33}$ is noticeably larger than $C_{11}$ and $C_{22}$ which indicates that the crystal is more resistant to deformation along the *c*-axis. This suggests stronger bonding between atoms in this direction and therefore a larger external force is required to distort the crystal structure along the *c*-axis.

The elastic constant $C_{44}$ represents the resistance of a crystal to shear deformation when a tangential force is applied on the (100) plane along the *b*-direction. Since a higher $C_{44}$ values are generally associated with greater hardness, HfRuP is expected to be harder than ZrRuAs. For both compounds, $C_{44}$ is smaller than $C_{11}$ and $C_{33}$. This indicates that shear deformation occurs more easily than uniaxial deformation along the principal crystallographic axes. The tetragonal shear modulus ($C_t$) reflects the crystal's rigidity and is closely related to the propagation of transverse acoustic waves and dynamical stability. The positive $C_t$ values obtained for both ZrRuAs and HfRuP suggest their dynamical stability [60].

The ductility or brittleness of a material can be predicted from the Cauchy pressure, defined as $C_P = (C_{12} - C_{44})$ [61]. A positive $C_P$ indicates ductile behavior, while a negative value suggests brittleness. The present study shows positive Cauchy pressure for both compounds. This indicates that they can undergo plastic deformation rather than fracturing suddenly under applied stress. The positive $C_P$ values also suggest a noticeable nondirectional (metallic, ionic) contribution to their bonding character whereas negative values are associated with covalent bonding [62,63].

The Kleinman parameter ($\zeta$) is a dimensionless measure of a crystal's internal response to deformation. It quantifies the relative displacement of cation and anion sublattices under volume-preserving strain when internal atomic coordinates are free to relax. This parameter provides insight into how the lattice adjusts internally during deformation [64]. The calculated $\zeta$ values are 0.53 for HfRuP and 0.52 for ZrRuAs. Since these values are close to the mid-range, the mechanical response of both compounds is expected to be governed by both bond-angle distortion and by direct bond stretching or compression. It can be computed as:

$$\zeta = \frac{C_{11} + 8C_{12}}{7C_{11} + 2C_{12}} \quad (2)$$

**Table 3.** Bulk $B$ (GPa), shear $G$ (GPa), Young's modulus $Y$ (GPa), Kleinman parameter ($\zeta$) machinability index $\mu_M$, Pugh's ratio $k$ ($G/B$), Hardness $H_v$ (GPa) and Poisson's ratio ($\nu$) of ZrRuAs and HfRuP compounds.

| Compound | $B$ | $G$ | $Y$ | $G/B$ | $\nu$ | $\zeta$ | $\mu_M$ | $H_v$ | Ref. |
|---|---|---|---|---|---|---|---|---|---|
| ZrRuAs | 200.30 | 95.33 | 246.84 | 0.48 | 0.29 | 0.52 | 2.01 | 13.39 | This work |
| HfRuP | 238.02 | 101.93 | 267.60 | 0.43 | 0.31 | 0.53 | 2.07 | 12.94 | This work |

The macroscopic mechanical properties listed in **Table 3** are derived from the elastic constants. These include the bulk modulus ($B$), shear modulus ($G$), Young's modulus ($Y$), hardness ($H$), Kleinman parameter ($\zeta$), machinability index ($\mu_M$), and Poisson's ratio ($\nu$). For a hexagonal lattice, the Voigt shear modulus ($G_V$) and Reuss shear modulus ($G_R$) are given by the well-known relations [65–67].

The bulk modulus indicates how strongly a material resists volume reduction under uniform pressure. Both the bulk modulus and shear modulus are associated with hardness. But, the shear modulus is generally regarded as a better qualitative indicator of a material's hardness [68]. According to Hill's approximation [65], the calculated shear modulus values for both compounds are much smaller than their corresponding bulk modulus values. This suggests that these materials are more easily deformed by shear stress than by compression. HfRuP shows higher bulk and shear moduli than ZrRuAs which implies that HfRuP possesses greater hardness.

Pugh's ratio [69] is widely used to distinguish ductile materials from brittle materials. A material is considered brittle when $G/B > 0.57$, whereas values below this threshold value indicate ductility. In the present study, the calculated $G/B$ values are 0.43 for HfRuP and 0.48 for ZrRuAs. These results confirm the ductile nature of both compounds.

Young's modulus ($Y$)and Poisson's ratio ($\nu$) were calculated from the evaluated bulk and shear moduli using the standard relations [70]:

$$Y = \frac{9BG}{3B + G} \; and \; \nu = \frac{3B - Y}{6B} \quad (3)$$

Young's modulus measures the stiffness of a material and represents the ratio of tensile stress to tensile strain [71]. It also provides insight into the resistance of a material to thermal shock. Higher values are often associated with a stronger covalent contribution to bonding [72]. As shown in **Table 3**, HfRuP has a higher Young's modulus than ZrRuAs which indicates that HfRuP is mechanically stiffer.

Poisson's ratio ($\nu$) is a useful indicator of shear stability, interatomic force character, and ductile-brittle behavior in solids [73]. For materials dominated by central interatomic forces, $\nu$ typically lies between 0.25 and 0.50 [74,75]. The calculated values for HfRuP and ZrRuAs are 0.31 and 0.29, respectively. Since both values exceed the critical limit of 0.26, these two compounds can be classified as ductile[73,76]. Moreover, $\nu$ is generally about 0.25 for ionic solids and around 0.10 for covalent solids [77]. The calculated values suggest ionic contribution with central interatomic interactions in both compounds.

The machinability index ($\mu_M = B/C_{44}$) is commonly used to evaluate the ease of machining, plastic response, and lubricating tendency of a material [78]. According to the formula, a smaller $C_{44}$ value results in a higher $\mu_M$ value, which indicates better machinability and lower friction during mechanical processing. The calculated $\mu_M$ values suggest that both compounds possess favorable machinability along with good lubricating characteristics.

Hardness refers to the ability of a material to oppose permanent deformation caused by actions such as indentation, scratching, wear, or penetration. For solid materials, a larger hardness value means that the material has stronger resistance against plastic deformation when an external force is applied. Therefore, the hardness of ZrRuAs and HfRuP is computed using $G$, $Y$, and B as follows [79,80]:

$$H_V = \frac{(1 - 2\nu)Y}{6(1 + \nu)} \quad (4)$$

Based on the calculated values of hardness, ZrRuAs exhibits a slightly higher hardness value than HfRuP, indicating its marginally greater resistance to localized plastic deformation. However, the low hardness values suggest that both compounds are mechanically soft and can undergo deformation relatively easily under an applied load. Although ZrRuAs is comparatively harder than HfRuP, both materials may be classified as soft, ductile, and machinable solids.

Elastic anisotropy plays a crucial role in crack propagation and the production of plastic deformation in solid materials. It affects a broad spectrum of physical phenomena, including the formation and spread of microcracks, crack motion, bonding characteristics across various crystallographic directions, the emergence of plastic deformations in solids, unconventional phonon modes, dislocation dynamics, internal abrasion, and more [81,82]. The universal anisotropy index, $A^U$ were evaluated for crystals of arbitrary symmetry using the standard relations given in **Refs.**[83–85]:

$$A^U = \frac{5G_V}{G_R} + \frac{B_V}{B_R} - 6 \quad (5)$$

The universal elastic anisotropy index ($A^U$) is zero for an elastically isotropic material while a non-zero value signifies elastic anisotropy. The calculated $A^U$ values for ZrRuAs and HfRuP are 0.04 and 0.19, respectively which indicates that HfRuP is moderately elastically anisotropic. On the other hand, ZrRuAs is almost isotropic.

The directional mechanical behavior of ZrRuAs and HfRuP was evaluated through 3D plots of

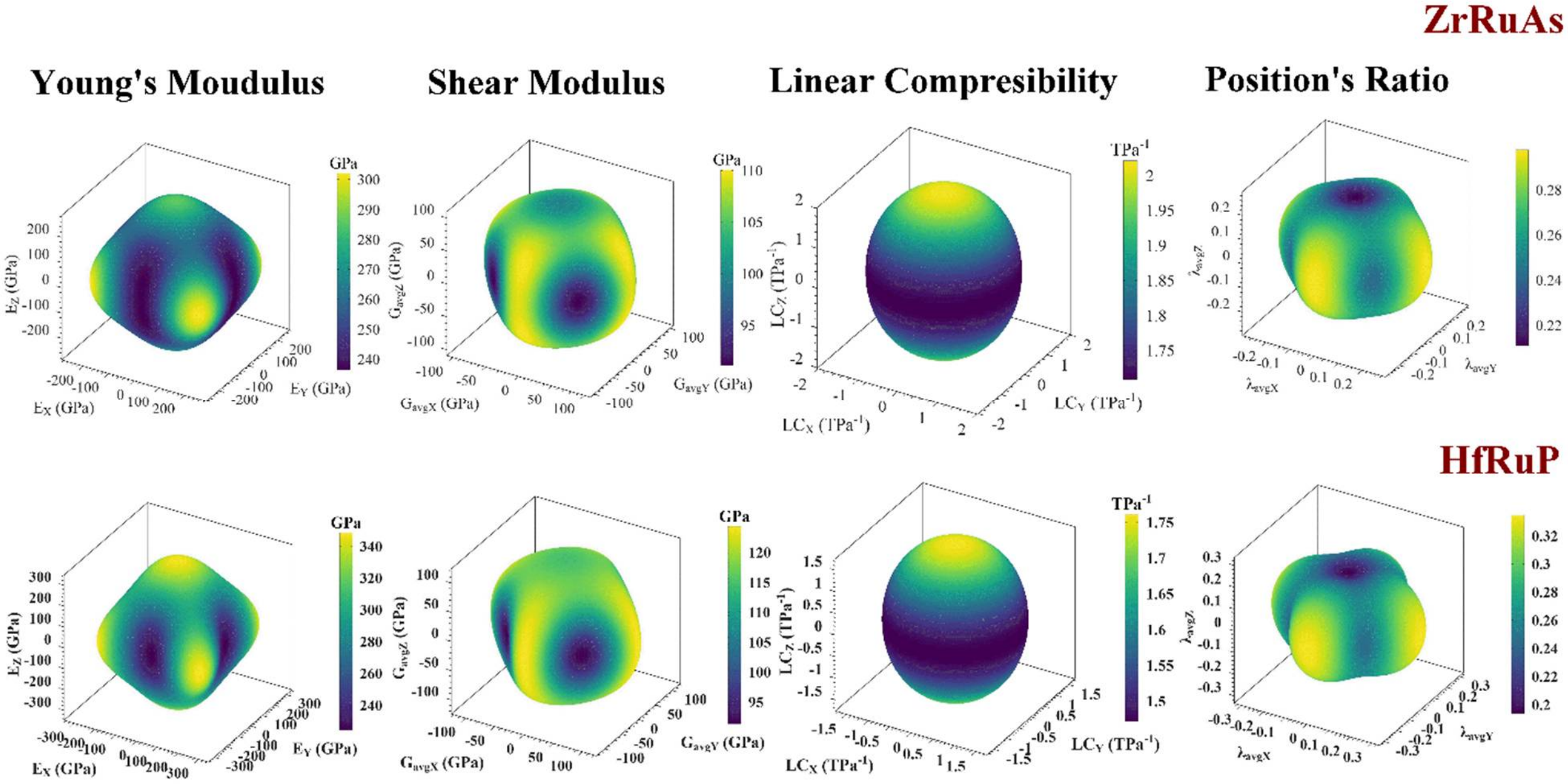

**Figure 3.** 3D representation of the mechanical properties of ZrRuAs and HfRuP in terms of Young's modulus, Shear modulus, Linear compressibility and Poisson's ratio.

Young's modulus, shear modulus, linear compressibility, and Poisson's ratio. The slightly non-spherical surfaces confirm weak elastic anisotropy in both compounds. Linear compressibility shows the most uniform directional response while Poisson's ratio exhibits the strongest anisotropy. The greater surface distortion of HfRuP indicates that it is more elastically anisotropic than ZrRuAs.

### 3.2.1 Electronic Band Structure

Electronic band structure is a fundamental tool for understanding the electronic properties of materials. The band structure is essential for investigating a material's properties, such as electrical and thermal conductivity, heat capacity, the Hall effect, and magnetic and optoelectronic behavior. The electronic band structure serves as a key framework for analyzing and understanding the intrinsic electronic behavior of materials. Examining the band structure is an indispensable step in predicting and interpreting the functional characteristics of a material [**86**–**88**]. Furthermore, band structure analysis identifies atoms and orbitals involved in electronic transition and probable band inversions associated with nontrivial topological phases [**89**,**90**]. Since both HfRuP and ZrRuAs crystallize in a hexagonal structure, their electronic band structures were computed along the high-symmetry path *Γ-M-K-Γ-A-L-H-A* in the Brillouin zone [**91**]. The calculated band structures without spin-orbit coupling (SOC) demonstrate that several electronic bands pass through the Fermi level ($E_F$) for both compounds, indicating a metallic or semi-metallic phase electronic character. In particular, up to four bands intersect $E_F$, with band crossings appearing along the *M-K*, *K-Γ* and *Γ-A* directions. Such band crossings near the Fermi level are characteristic of semi-metallic systems and are commonly observed in topological materials [**92**–**94**]. Furthermore, linear band dispersion with inversion is observed close to the *K* point in both compounds, which is a key signature associated with topologically nontrivial electronic states. According to the theory of topological quantum chemistry, the inversion of bands with different orbital characters at high-symmetry points can lead to Weyl semimetal, Dirac semimetal, or topological insulating phases [**89**,**90**].

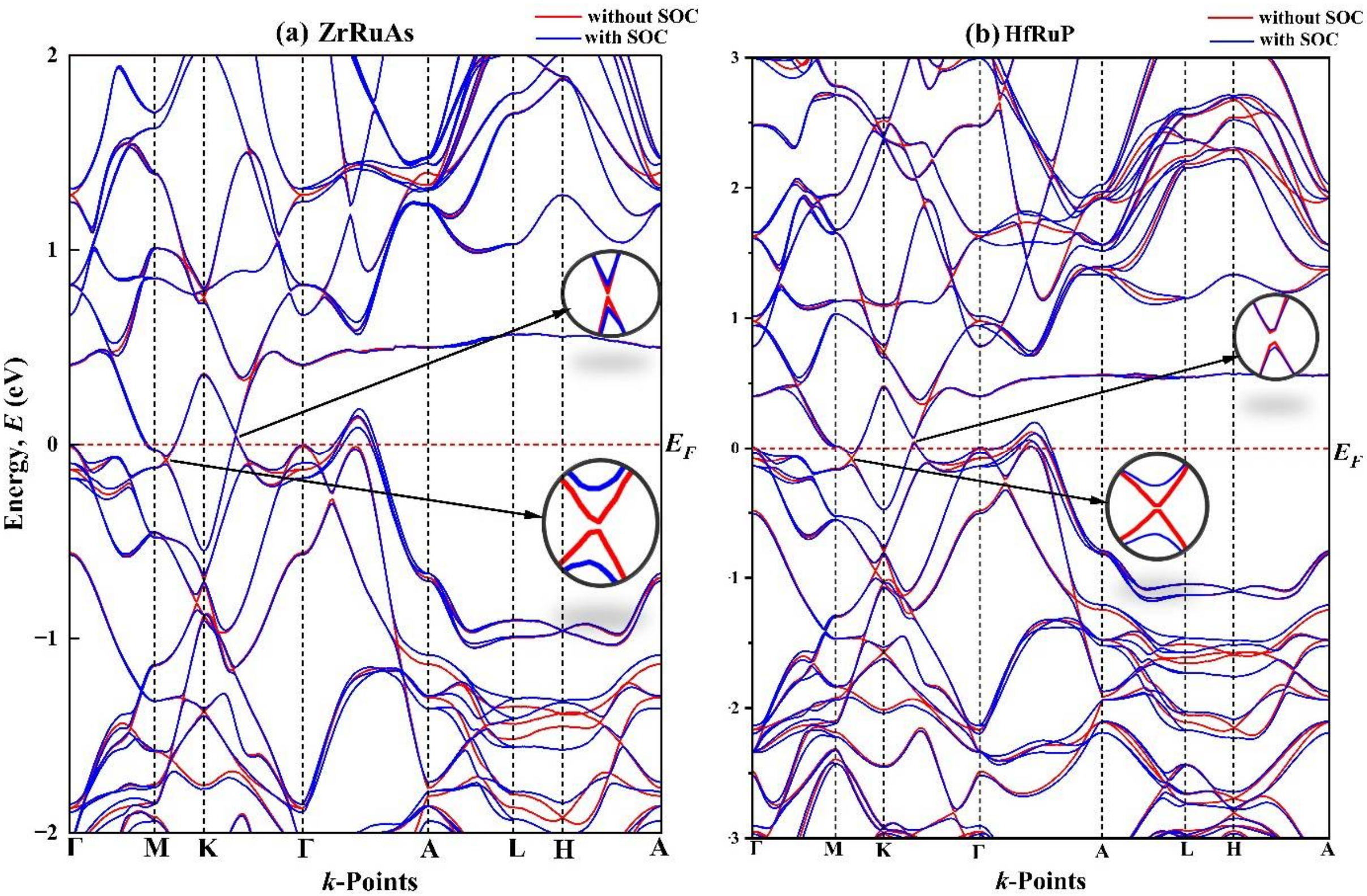

**Figure 4.** Calculated electronic band Structure of ZrRuAs and HfRuP without and with SOC.

To further elucidate relativistic effects and topology, the electronic band structures were computed including the SOC. Upon addition of SOC, the overall band dispersion stays qualitatively identical. However, several degeneracies observed in the non-SOC calculations are lifted, leading to evident band splitting. This effect is expected in non-centrosymmetric systems, where SOC reduces spin degeneracy through antisymmetric spin-orbit interactions [95,96]. The SOC-induced splitting is more apparent in HfRuP, indicating the higher relativistic effects related with the heavier Hf atom compared to Zr in ZrRuAs. In certain regions of the Brillouin zone, particularly along the *M*-*K* direction, SOC induces a band gap opening at the band crossing, resulting in local band gaps. Along the *K*-*Γ* direction, ZrRuAs exhibits a more distinct SOC-induced gap opening, whereas the band crossings in HfRuP remain largely robust, preserving a gapless semi-metallic character. Such SOC-induced band gap is commonly observed in nodal-line or band-inverted systems when mirror or inversion symmetries are partially lifted [97–99]. While SOC can open a full bulk gap in some systems, it may also gap nodal lines everywhere except at isolated points, giving rise to Weyl or Dirac nodes [97–100].

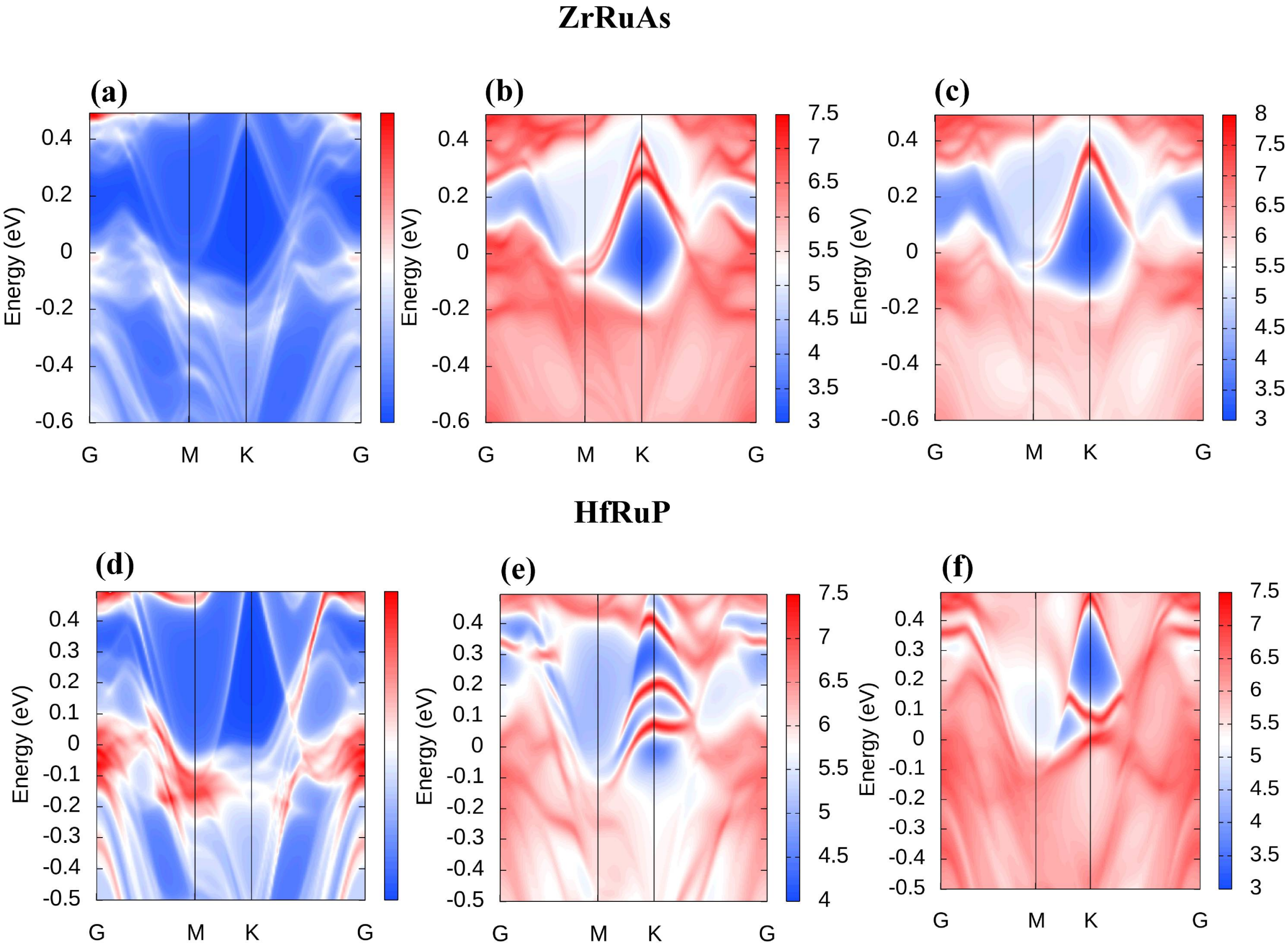

**Figure 5.** Bulk, top and bottom surface density of states of ZrRuAs (a-c) and HfRuP (d-f), respectively. The color scale represents the spectral intensity near the Fermi level.

The persistence of gapless band crossings in HfRuP under SOC is consistent with previous theoretical work identifying this compound as a Weyl semimetal hosting twelve pairs of type-II Weyl points [58]. In contrast, ZrRuAs has been classified as a topological crystalline insulator characterized by trivial Fu-Kane $Z_2$ indices but nontrivial mirror Chern numbers [58]. Although the present bulk band structure calculations display semi-metallic features near the Fermi level, such behavior is compatible with a topological crystalline insulating classification, as small band overlaps are frequently observed in density functional theory calculations of topological insulators due to exchange-correlation approximations [101]. The surface spectral functions obtained from the MLWF-derived tight-binding model reveal distinct surface features near the $K$ point (see **Figure 5**). In ZrRuAs, the surface states exhibit partially gapped dispersions around the Fermi level which is consistent with the expected behavior of a topological crystalline insulator. In contrast, HfRuP shows more robust gapless surface states near the Fermi level that supports its Weyl semimetal character [22,23].

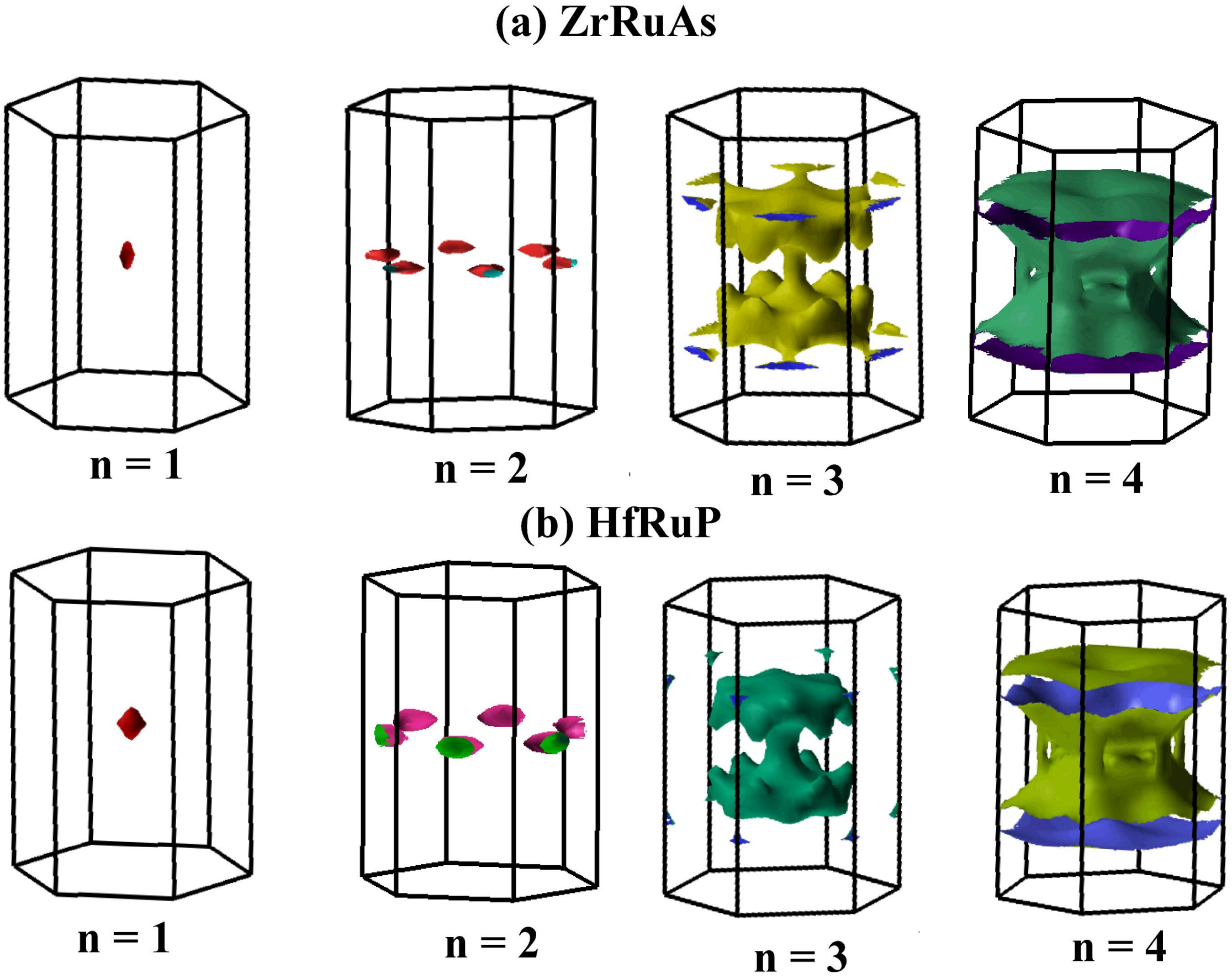

**Figure 6.** Fermi surface topologies of (a) ZrRuAs and (b) HfRuP.

For hexagonal ZrRuAs and HfRuP, four dispersive electronic bands cross the Fermi level. This produces four distinct Fermi surface sheets, as shown in **Figure 6(a) and (b).** The small Fermi surface sheet (n =1) in **Figure 6(a)** exhibits hole-like character and encloses the $\Gamma$ high-symmetry point. The Fermi surface sheets (n = 2, 3) shown in **Figure 6(a)** form multiple connected and tubular sheets around the $\Gamma$ point. These sheets predominantly show hole-like behavior with possible mixed character in certain regions. In contrast, the Fermi surface sheet (n = 4) in **Figure 6(a)** exhibits electron-like character and forms a closed curtain-like structure.

In general, HfRuP shows a Fermi surface topology similar to that of ZrRuAs which indicates comparable electronic behavior near the Fermi level. However, a noticeable difference is observed in the third Fermi surface sheet (n = 3) in **Figure 6(b)**. In ZrRuAs, this sheet forms a more extended and nearly continuous hollow-like topology accompanied by small additional pockets near the Brillouin zone boundary. In contrast, the corresponding sheet in HfRuP appears comparatively compact, distorted, and more localized around the central region of the Brillouin zone. This difference suggests relatively stronger band dispersion near the Fermi level in ZrRuAs compared with HfRuP.

Overall, the band structure analysis demonstrates that SOC plays a crucial role in lifting band degeneracies and modifying band crossings in both compounds. The results are consistent with the distinct topological nature of HfRuP as a type-II Weyl semimetal and ZrRuAs as a topological crystalline insulator, in agreement with earlier theoretical predictions [**22**,**23**].

### 3.2.2 Density of States

The total and atomic orbital-projected density of states (DOS) of ZrRuAs and HfRuP were calculated to elucidate the nature of the electronic states around the Fermi level ($E_F$ = 0 eV), as shown in **Figure 7**. The DOS is resolved into atomic and orbital contributions to identify the dominant states governing the low-energy electronic behavior of these compounds. The total DOS of ZrRuAs exhibits a reduction in spectral weight at the Fermi level that indicates a semimetallic character. The orbital-resolved DOS reveals that the Ru 4*d* states dominate the valence band from approximately -6 eV up to $E_F$, while their contribution rapidly decreases near the Fermi level. The As 4*p* states are mainly localized in the valence band (-6 to -2 eV) where they strongly hybridize with Ru 4*d* orbitals. This indicates significant covalent bonding between Ru and As atoms. Contributions from Zr 4*d* states are primarily located above the Fermi level in the conduction band with only negligible weight at $E_F$ and the Zr 5*s* and 4*p* states contribute minimally near the Fermi energy. The suppression of bulk DOS at $E_F$, together with strong Ru *d* and As *p* hybridization away from the Fermi level, suggests a reduced density of bulk carriers in ZrRuAs. Such an electronic structure may be compatible with previously reported topological characteristics in related compounds where the bulk electronic states are weakly populated near $E_F$ while metallic states can emerge at symmetry-protected surfaces.

Moreover, the total DOS of HfRuP shows a finite and appreciable density of states at the Fermi level which indicates a metallic or semimetallic ground state. The projected DOS further reveals that Hf *5d* states contribute significantly near and above the Fermi level which reflects their more extended character and stronger relativistic effects. The Ru 4*d* orbitals dominate the valence band region and retain appreciable weight at $E_F$ indicating their important role in low-energy electronic transport. In contrast, the P 3*p* states are mainly confined to the lower valence band (-6 to -3 eV) and contribute weakly near the Fermi level through hybridization with Ru *d* states. The coexistence of Hf 5*d* and Ru 4*d* states at the Fermi energy, together with the stronger spin-orbit coupling (SOC) associated with Hf, may promote band crossings in the vicinity of $E_F$. In non-centrosymmetric systems such crossings are not always fully gapped by SOC, which can potentially lead to Weyl nodes in the bulk band structure. The finite DOS at $E_F$ is therefore consistent with the semimetallic electronic character reported for HfRuP.

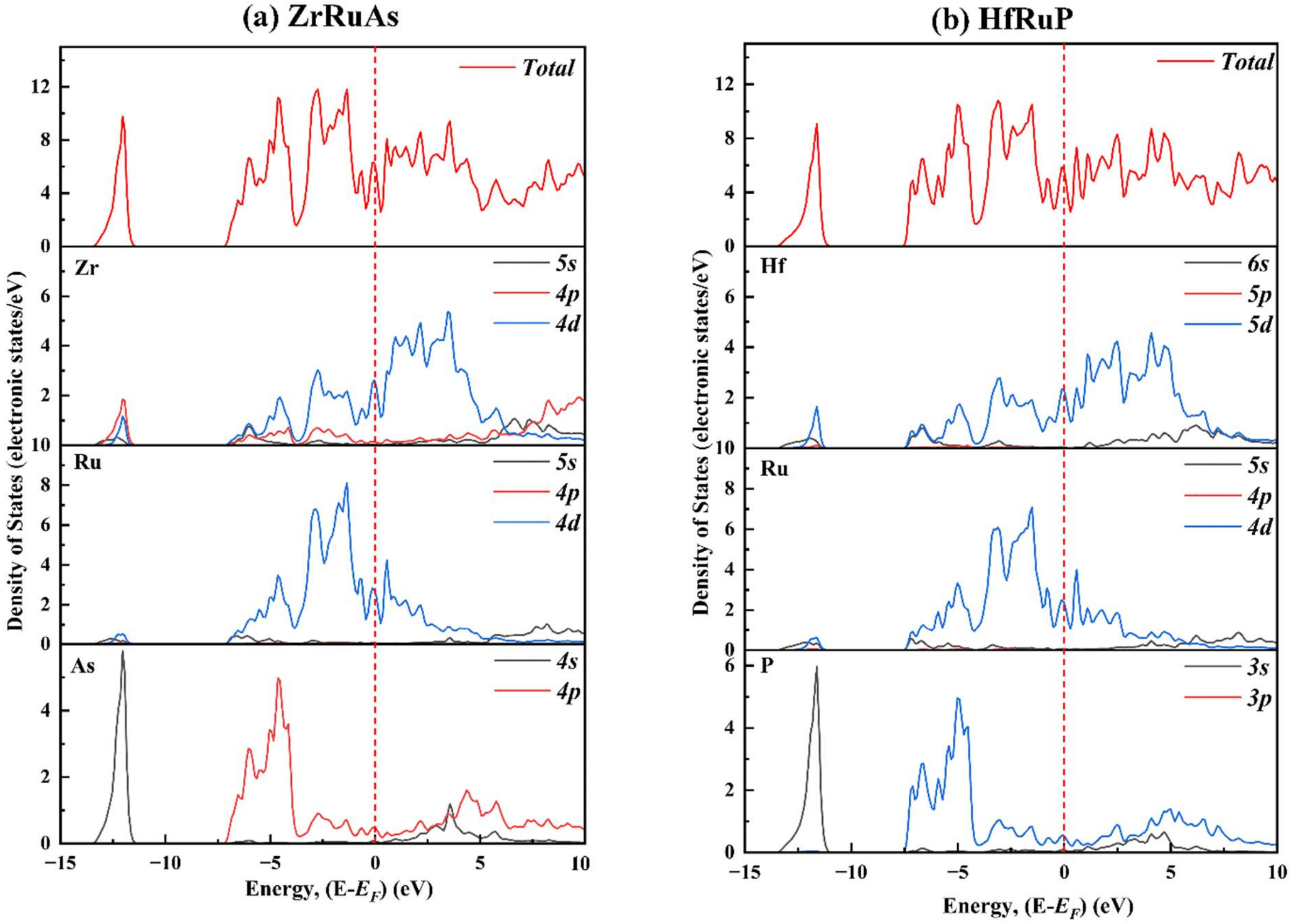

**Figure 7.** The electronic total and partial density of states of (a) ZrRuAs and (b) HfRuP.

## 3.3 Optical Properties

### 3.3.1 Dielectric Function

The dielectric function is a fundamental parameter for analyzing a material's optical behavior because it describes how light interacts with the system. Specifically, how the material responds to an external electromagnetic field. The real component of dielectric function primarily represents polarization and energy storage within the material. In contrast, the imaginary component is associated with optical absorption caused by electronic transitions. These components are used to determine several important optical quantities including the refractive index, extinction coefficient, absorption coefficient, and reflectivity. The total dielectric response generally arises from both intraband and interband transitions. In metallic systems, the interband contribution is particularly important for describing the optical spectra at low photon energies [102]. Interband transitions are further categorized into direct and indirect transitions. In the present analysis, indirect interband transitions are neglected because they involve phonon-assisted scattering and are generally expected to provide only a minor contribution to $\varepsilon(\omega)$. Because both the compounds are metallic, a Drude contribution was included using a plasma energy of 10 eV, damping of 0.05 eV for ZrRuAs and plasma energy of 5 eV, damping of 0.05 eV for HfRuP [103,104]. For anisotropic materials, the dielectric function is expressed as a complex symmetric second-order tensor that describes the linear response of the system to an external electric field. To examine the optical anisotropy, the spectra were calculated for electric-field polarizations along the [100] and [001] directions over the photon energy range up to 30 eV for both materials. As shown in **Equation 6**, the dielectric function consists of real and imaginary parts. The real part was derived from the imaginary component using the Kramers-Kronig relation given in **Equation 7**,

$$\varepsilon(\omega) = \varepsilon_1(\omega) + i\varepsilon_2(\omega) \quad (6)$$

$$\varepsilon_1(\omega) = 1 + \frac{2}{\pi} P \int_0^\infty \frac{\omega' \varepsilon_2(\omega')}{\omega'^2 - \omega^2} d\omega' \quad (7)$$

where $P$ implies the principal value of the integral.

Large negative $\varepsilon_1(\omega)$ at low energy indicates Drude-like metallic response **Figure 8(a)**. In **Figure 8(b)** $\varepsilon_2(\omega)$ approaches zero near ~25 eV for ZrRuAs and ~27 eV for HfRuP, implying transparency above this energy. The effective plasma frequency $\omega_P$ lies near ~25 eV for ZrRuAs and ~27 eV for HfRuP for both polarizations.

### 3.3.2 Refractive Index and Reflectivity

The refractive index is a complex quantity, defined as $\tilde{n}(\omega) = n(\omega)+ik(\omega)$, where $n(\omega)$ represents the real part controlling the speed of light propagation in the material and $k(\omega)$ represents the extinction coefficient quantifying the attenuation of light due to absorption. Larger values of $k(\omega)$ indicate stronger optical absorption which is a key parameter in the design of photonic and optoelectronic devices. From **Figure 8(e)**, $n(\omega)$ attains its maximum at zero photon energy and gradually decreases with increasing photon energy. Both HfRuP and ZrRuAs maintain relatively high refractive indices across the visible spectrum, highlighting their potential for applications in optoelectronic devices such as display technologies, where efficient light manipulation is required.

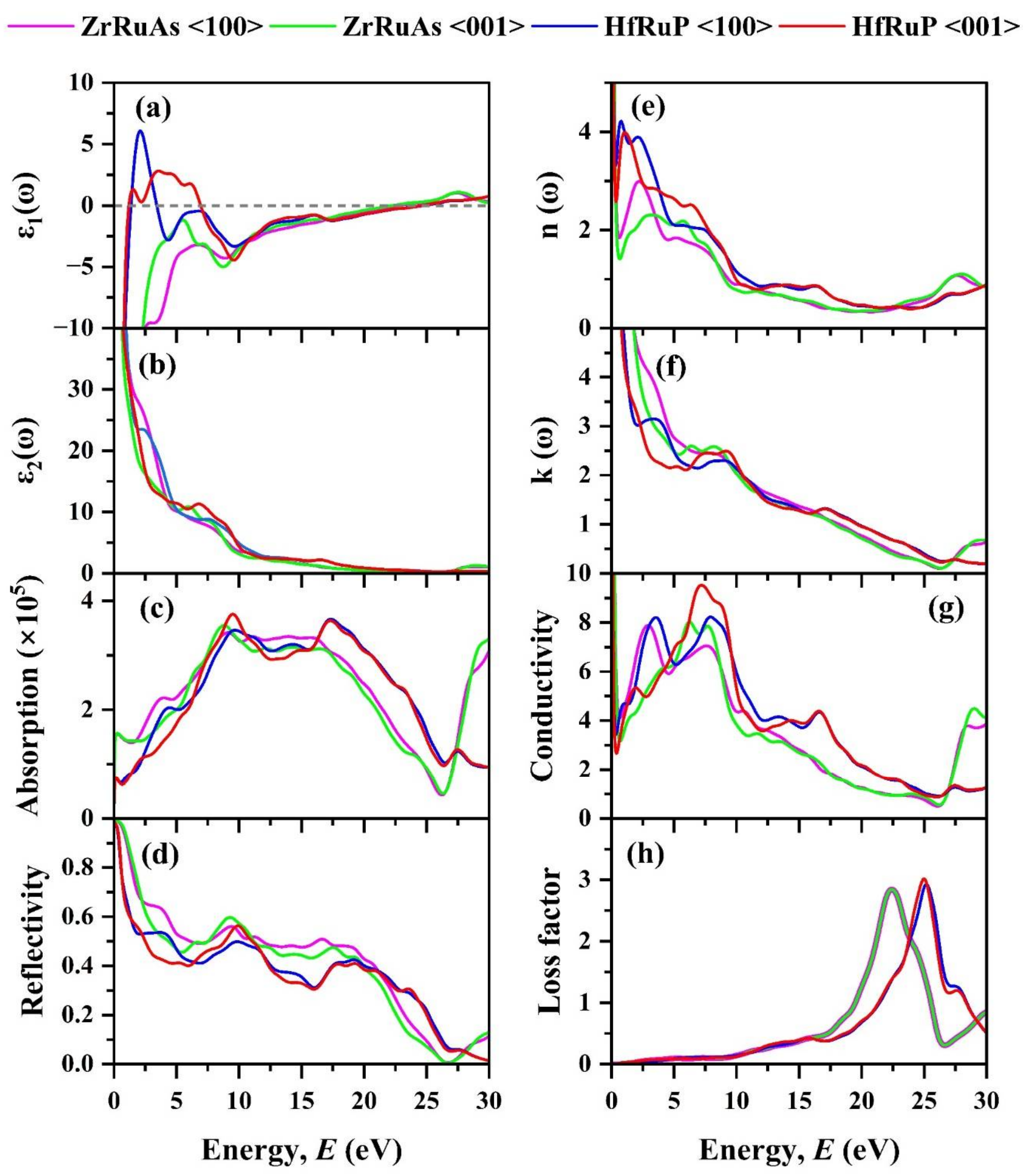

**Figure 8.** The energy dependent (a) real part, $\varepsilon_1$, (c) imaginary part, $\varepsilon_2$ of dielectric function, (b) refractive index, $n(\omega)$, (d) extinction coefficient, $k(\omega)$, (e) absorption coefficient, $\alpha(\omega)$, (f) optical conductivity, $\sigma_o(\omega)$, (g) loss function, $L_0(\omega)$ of ZrRuAs and HfRuP with electric field polarization vectors along [100] and [001] directions.

Reflectivity is an important parameter for probing the surface optical response and understanding light-matter interactions in a material. It is particularly sensitive near electronic resonance regions where variations in reflectivity can reveal interband transitions and underlying features of the electronic band structure [105-109]. The reflection coefficient $R(\omega)$ depends on the energy of the incident photons which quantifies the fraction of electromagnetic energy reflected at the material's surface. This coefficient is defined according to **Equation 8** [110]:

$$R(\omega) = \left|\frac{\sqrt{\varepsilon(\omega)} - 1}{\sqrt{\varepsilon(\omega)} + 1}\right|^2 = \left|\frac{\sqrt{\varepsilon_1(\omega) + i\varepsilon_2(\omega)} - 1}{\sqrt{\varepsilon_1(\omega) + i\varepsilon_2(\omega)} + 1}\right|^2 \quad (8)$$

Optical reflectivity is highest at zero photon energy for both compounds and decreases almost linearly up to approximately 2.5 eV which indicates a free-electron like metallic response [see **Figure 8(d)**]. The maximum reflectivity at zero photon energy confirms the compounds shiny, metallic appearance. Moreover, $R(\omega)$ remains above 40% up to about 22 eV for HfRuAs and above 50% up to 22 eV for ZrRuP which suggests strong solar reflectance in this spectral range.

### 3.3.3 Absorption and Conductivity

Light absorption and optical conductivity are key processes that describe the interaction of electromagnetic radiation with a material. During absorption, the energy of incident photons is transferred to the system primarily through electronic transitions. This provides valuable insights into the electronic structure and energy levels of the material. Optical conductivity measures the material's capability to produce an electric current in response to incident light and depends both on the intrinsic electronic properties of the material and the frequency of the applied radiation. The optical absorption coefficient $\alpha(\omega)$ can be derived from the real and imaginary components of the dielectric function [110].

$$\alpha(\omega) = \frac{\sqrt{2}\omega}{c}\left[\sqrt{\varepsilon_1(\omega)^2 + \varepsilon_2(\omega)^2} - \varepsilon_1(\omega)\right]^{1/2} = \frac{4\pi k(\omega)}{\lambda} \quad (9)$$

Optical absorption arises from both interband electronic transitions and intraband (Drude) contributions [111]. The positions of the peaks in the absorption spectrum correspond directly to the maxima in the imaginary component of the dielectric function, $\varepsilon_2(\omega)$. The absorption observed in the low energy region (0-20 eV) can be attributed to transitions between closely spaced electronic states within the bands which results in a broadened absorption feature [see **Figure 8(c)**]. In contrast, at higher energies the absorption mainly arises from transitions between the valence and conduction bands. These transitions are more well defined which produces much sharper spectrum. For both compounds, optical absorption begins from zero photon energy which is consistent with their metallic electronic character. This low energy absorption behavior also suggests that these materials may be relevant for light harvesting and solar-energy related applications. As shown in the absorption spectra, the strongest absorption response is mainly observed within the energy interval of approximately 9-17 eV for both HfRuP and ZrRuAs. Beyond this region, it drops rapidly near 20 eV, which corresponds well with the occurrence of prominent peaks in the energy-loss spectra.

The optical conductivity spectrum can be presented in range of energies between 0-30 eV, using the imaginary part of the dielectric function and the equation [110]:

$$\sigma(\omega) = \frac{\omega}{4\pi}\varepsilon_2(\omega) \tag{10}$$

In both compounds, $\sigma(\omega)$ is non zero at zero photon energy, implying there is no band gap and the materials behave metallically. As photon energy increases, the photoconductivity grows to a peak, then gradually declines at higher energies and eventually tends toward zero.

### 3.3.4 Loss Spectrum

The energy-loss function $L(\omega)$, is defined as the inverse of the imaginary part of the dielectric function and is commonly expressed as [110]:

$$L(\omega) = \mathrm{Im}\left[-\frac{1}{\varepsilon(\omega)}\right] \tag{11}$$

The energy-loss spectrum offers valuable information about both the macroscopic and microscopic electronic properties of a material. The most prominent feature in this spectrum is the plasmon peak, which arises from the collective oscillation of valence electrons and occurs at the plasma frequency. As illustrated in **Figure 8(h),** the position of this peak indicates the energy at which the material changes from metallic to dielectric behavior. Below the plasma energy, the compound exhibits metallic characteristics whereas the dielectric features become dominant above it [112]. The plasmon peak occurs at ~23 eV for ZrRuAs and ~26 eV for HfRuP. Higher plasmon peak in HfRuP indicates that free electron concentration is higher in this compound compared to ZrRuAs.

## 3.4 Thermodynamic properties

The thermophysical properties of the Ru-based ternary compounds ZrRuAs and HfRuP were examined under varying pressures and temperatures using first-principles calculations within the framework of the quasi-harmonic Debye approximation. The isothermal-isobaric thermodynamic properties were evaluated using the Gibbs2 program, where the relevant thermodynamic parameters were derived from the calculated energy-volume relationships based on the quasi-harmonic Debye model [50]. Within this framework, the non-equilibrium Gibbs function $G^*(V; P, T)$ can be expressed as follows,

$$G^*(V;P,T) = E(V) + PV + A_{vib}[\Theta_D(V);T] \tag{12}$$

where, $E(V)$ denotes the total energy for unit cell, $PV$ represents the contribution arising from the applied hydrostatic pressure, $\Theta_D(V)$ is the Debye temperature and $A_{\text{vib}}$ is the vibrational Helmholtz free energy. Within the model of the phonon density of states $A_{\text{vib}}$ can be expressed as follows:

$$A_{vib}[\Theta_D(V);T] = nk_BT\left[\frac{9\Theta_D}{8T} + 3ln\left(1 - e^{-\frac{\Theta_D}{T}}\right) - D\left(\frac{\Theta_D}{T}\right)\right] \tag{13}$$

where, $n$ represents the number of atoms per formula unit, $k_B$ is the Boltzmann constant, $D\left(\frac{\Theta_D}{T}\right)$ denotes the Debye integral, and $\Theta_D$ is the Debye temperature for an isotropic solid [113].

The Debye temperature ($\Theta_D$) is an important thermodynamic parameter because it is closely related to several fundamental physical properties, including specific heat capacity, thermal conductivity, electrical transport behavior, and thermal expansion. The temperature dependence of ($\Theta_D$) for ZrRuAs and HfRuP at pressures of 0, 10, 20, 30, 40, and 50 GPa is presented in **Figure 9(c) and (d)**, respectively. The Debye temperature exhibits a standard decreasing trend with increasing temperature in both compounds. The application of external pressure leads to a noticeable increase in its value. At low temperatures, the Debye temperature tends to approach a nearly plateau-like region. This behavior is commonly associated with zero-point lattice vibrations, which remain present even in the limit of very low temperatures [**114**,**115**]. The Debye temperatures were also estimated using the Anderson approach [116] based on the calculated elastic constants. The computed values are 424 K for ZrRuAs and 401 K for HfRuP. These values are in good agreement with those predicted by the quasi-harmonic Debye model. Because $\Theta_D$ is directly influenced by the average velocity of sound waves propagating through a crystal and these sound velocities can be determined by the elastic constants. The consistency between the two approaches highlights the strong coupling between the elastic response and lattice vibrational characteristics of these materials [**116**]. A thermodynamic quantity that describes the resistance to compression is the isothermal bulk modulus. **Figure 9(a) and (b)** illustrate the variation of bulk modulus ($B$) with temperature (0-1000 K) at different pressures for ZrRuAs and HfRuP. For both compounds, $B$ decreases monotonically with increasing temperature at all applied pressures. HfRuP exhibits consistently higher bulk modulus values compared to ZrRuAs which suggests stronger interatomic bonding and greater mechanical rigidity [**117**].

**Figure 9(e)-(h)** present the temperature dependence of heat capacities at constant volume ($C_V$) and constant pressure ($C_P$) for ZrRuAs and HfRuP. At low temperatures, $C_V$ increases rapidly which is consistent with the Debye-$T^3$ power law. At higher temperatures ($T$ > 600 K), $C_V$ approaches the classical Dulong-Petit limit (225 $\text{Jmole}^{-1}\text{K}^{-1}$), confirming the validity of the quasi-harmonic approximation used in similar thermophysical studies [**117**,**118**]. The same trend is observed for specific heat at constant pressure ($C_P$) and is shown in **Figure 9(g) and (h)**. The temperature dependence of $C_P$ for ZrRuAs and HfRuP exhibits the characteristic phonon-driven behavior of crystalline solids, with a rapid increase at low temperature followed by saturation near the Dulong-Petit limit at elevated temperature. The slight decrease in $C_P$ with increasing pressure indicates lattice stiffening and enhancement of Debye temperature [**119**]. No anomalous behavior is observed, confirming thermodynamic stability of both compounds within the investigated temperature and pressure range.

**Figure 10(a)-(d)** depict the variation of thermal expansion coefficient (α) for ZrRuAs and HfRuP at 0-1000 K and 0-50 GPa pressure. As illustrated in **Figure 10(a) and (b),** under a fixed pressure, the thermal expansion coefficient ($\alpha$) increases sharply with temperature below 400 K for both compounds which indicates strong lattice expansion in the low-temperature region. Above 400 K, the increase in thermal expansion coefficient ($\alpha$) becomes more gradual that suggests reduced temperature sensitivity at higher temperatures.

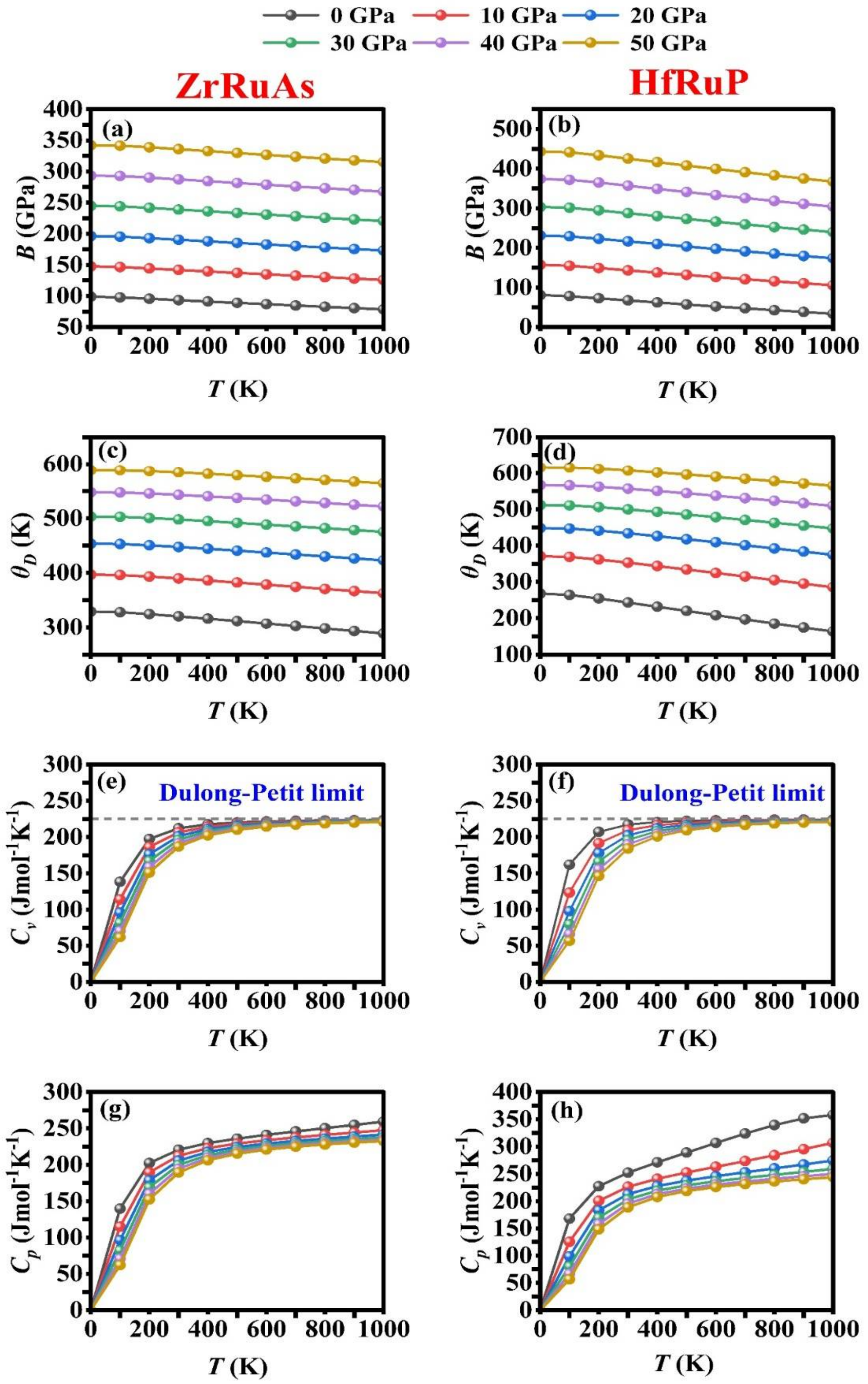

**Figure 9.** Pressure and temperature dependences of the thermodynamic properties including Bulk modulus ($B$), Debye temperature ($\Theta_D$), and heat capacity ($C_p$, $C_v$) for ZrRuAs (a, c, e, f) and HfRuP (b, d, f, h) compounds.

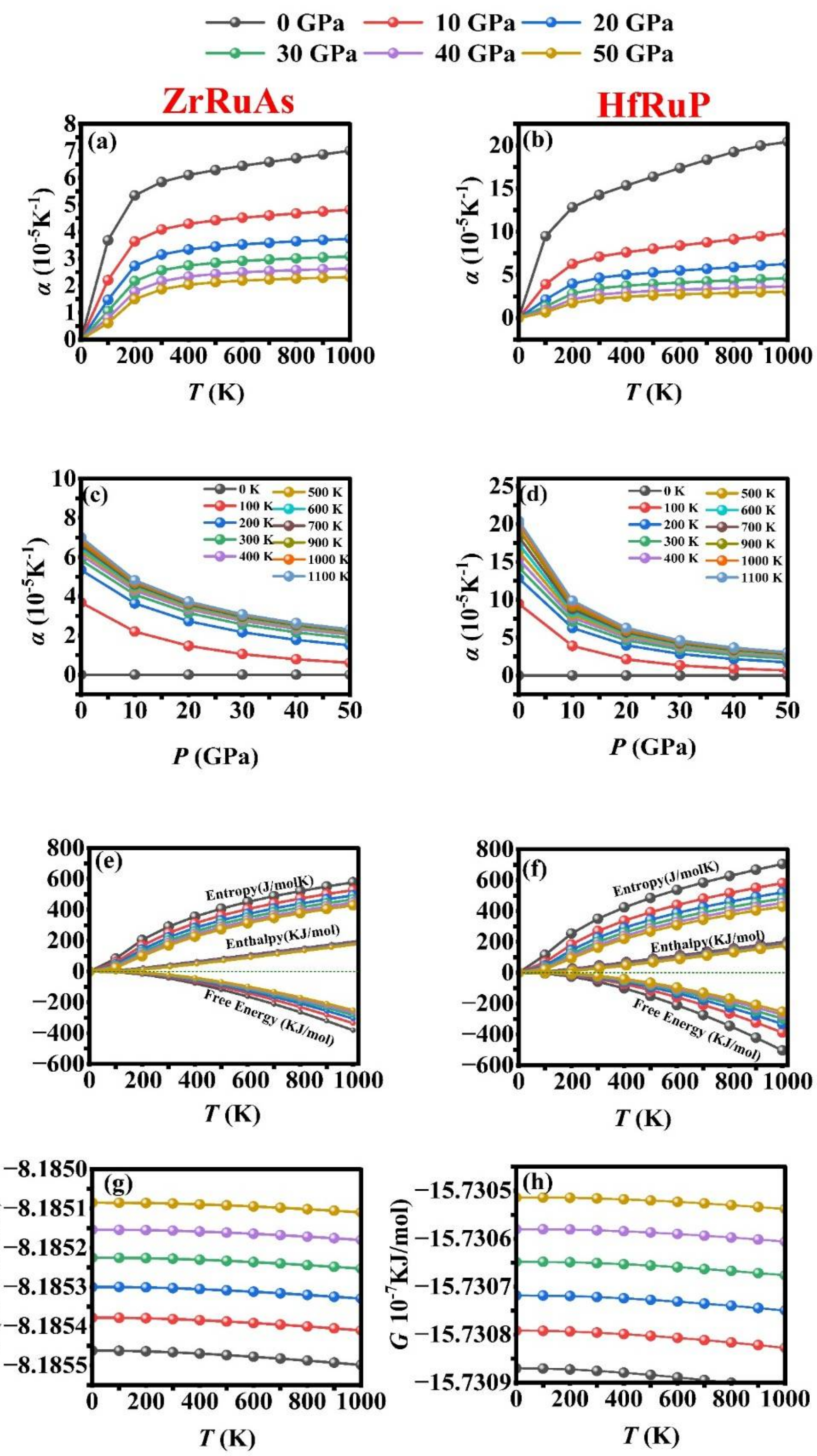

**Figure 10.** Pressure and temperature dependences of the thermodynamic properties including Expansion coefficient (α), Gibbs free energy (*G*), Entropy, Enthalpy and Free Energy of ZrRuAs (a, c, e, f) and HfRuP (b, d, f, h) compounds.

In addition, at constant temperature, the thermal expansion coefficient decreases with increasing pressure which indicates suppression of lattice expansion under compression. The pressure effect is more pronounced at lower pressures, as reflected by the steeper slopes. It gradually weakens at higher pressures. These results confirm that the thermal expansion behavior of ZrRuAs and HfRuP is governed by the combined influence of temperature and pressure [**120**]. The variation of Gibbs free energy $G(T, P)$ with temperature (0-1000 K) under pressures ranging from 0 to 50 GPa for ZrRuAs and HfRuP is presented in **Figure 10(g) and (h)**. Gibbs free energy is expressed as $\Delta G = \Delta H - T\Delta S$, where $\Delta H$ and $\Delta S$ represents the enthalpy and entropy changes of the system. As shown in **Figure 10(g)**, Gibbs free energy decreases progressively with increasing temperature. A comparative analysis reveals that HfRuP generally exhibits lower Gibbs free energy values than ZrRuAs at corresponding temperatures and pressures. Lower Gibbs free energy indicates enhanced thermodynamic stability [**118**].

Furthermore, the thermodynamic properties of ZrRuAs and HfRuP can be characterized using temperature-dependent thermodynamic potentials, including the enthalpy $H(T)$, Helmholtz free energy $F(T)$, and entropy $S(T)$. The variation of enthalpy, free energy and entropy with temperature for both compounds is shown in **Figure 10(e) and (f).** The calculated entropy of ZrRuAs and HfRuP increases monotonically with temperature in the range 0-1000 K. This behavior is typical for crystalline solids and originates from enhanced phonon population with increasing thermal excitation [**108**,**121**]. The enthalpy increases smoothly and nearly linearly with temperature for both compounds. This behavior confirms the absence of any structural phase transition within the investigated temperature range [**122**,**123**]. The Helmholtz free energy decreases continuously with increasing temperature at different pressures for both compounds. HfRuP demonstrates comparatively enhanced thermodynamic stability due to its more negative free energy values [**123**].

## 3.5 Superconducting Properties

**Figure 11(a) and (c)** presents the phonon dispersion curves and projected phonon density of states (PhDOS) of ZrRuAs and HfRuP, respectively. The absence of imaginary phonon frequencies throughout the Brillouin zone confirms the dynamical stability of both compounds. In the low-frequency region, the phonon modes are predominantly contributed by the out-of-plane vibrations of Ru atoms in both materials. The mid-frequency region mainly originates from mixed vibrations of Zr/Hf and Ru atoms, with partial contributions from As/P atoms. In the high-frequency region, the phonon modes are dominated by the lighter pnictogen atoms (As and P), as clearly demonstrated in the projected PhDOS. A notable difference between the two compounds is the presence of distinct optical phonon gaps in HfRuP, particularly around 140-200 $cm^{-1}$ and 260-330 $cm^{-1}$, whereas ZrRuAs exhibits a comparatively continuous phonon spectrum without a pronounced optical gap. This behavior indicates relatively stronger phonon mode separation in HfRuP. The electron-phonon coupling (EPC) weighted phonon dispersions together with the Eliashberg spectral functions $\alpha^2F(\omega)$ and integrated EPC constants [$\lambda(\omega)$] are shown in **Figure 11(b) and (d)**. In both compounds, the strongest EPC originates from softened low-frequency phonon modes near the $M$ and $K$ symmetry points, mainly associated with out-of-plane Ru vibrations. The most significant contribution to the total EPC originates from phonons below 100 $cm^{-1}$, contributing approximately 78.19% and 68.03% of the total EPC for ZrRuAs and HfRuP, respectively. Although $\alpha^2F(\omega)$ show noticeable spectral weight in the high-frequency region (above 100 cm$^{-1}$), its contribution to the total EPC remains limited. Such phenomena can be easily understood from **Equation 15**.

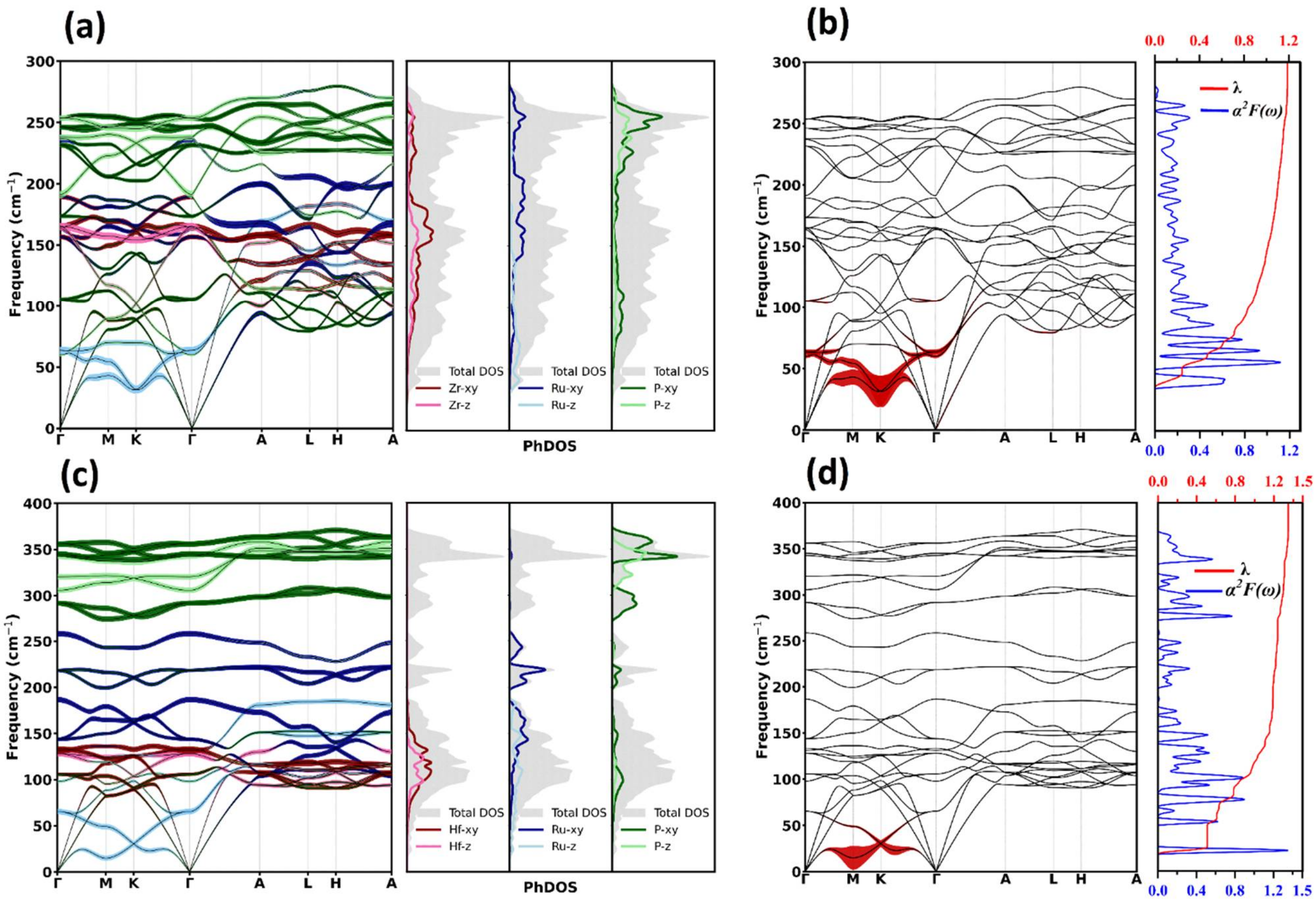

**Figure 11.** Phonon dispersions spectra of (a) ZrRuAs and (c) HfRuP weighted by different atomic vibrational modes. The right-side panel present the total phonon density of states (gray area) together with the mode-resolved PhDos curves (colored lines). The electron-phonon coupling (EPC) strength weighted phonon dispersions with the Eliashberg spectral function, $\alpha^2F(\omega)$ and the integrated strength of EPC, $\lambda(\omega)$ for (b) ZrRuAs and (c) HfRuP.

**Table 4.** Calculated superconducting parameters for ZrRuAs and HfRuP, including the Fermi level electronic energy density of states, $N(E_F)$ (in states/eV. unit cell), the electron-phonon coupling constant ($\lambda$), the logarithmic phonon frequency ($\omega_{log}$) and superconducting transition temperature, $T_c$ (K).

| Compound | $N(E_F)$ | $\lambda$ | $\omega_{log}$ | $T_c$ |
|---|---|---|---|---|
| ZrRuAs | 2.50 | 1.18 | 111.736 | 9.751 |
| HfRuP | 2.35 | 1.30 | 91.506 | 9.025 |

The superconducting behavior of ZrRuAs and HfRuP was investigated through electron-phonon coupling (EPC) calculations. The superconducting transition temperature ($T_c$) was then estimated using the Allen-Dynes modified McMillan equation. [124] given by:

$$T_c = \frac{\omega_{log}}{1.2} exp\left[\frac{-1.04(1+\lambda)}{\lambda - \mu^*(1+0.62\lambda)}\right] \quad (14)$$

Here, $\omega_{\log}$ represents the logarithmically averaged characteristic phonon frequency and is expressed as:

$$\omega_{log} = \exp\left[\frac{2}{\lambda}\int \frac{d\omega}{\omega} \alpha^2 F(\omega) \ln \omega\right] \quad (15)$$

In **Equation 14**, the $\mu^*$ represents the effective repulsive Coulomb pseudopotential. In this present work, $\mu^* = 0.10$ was adopted which is consistent with the widely accepted typical range between 0.10 and 0.16 [**125**-**127**]. The integrated EPC constant can be expressed as:

$$\lambda(\omega) = 2\int_0^\infty \frac{\alpha^2 F(\omega)}{\omega} d\omega \quad (16)$$

The Eliashberg electron-phonon spectral function, $\alpha^2$F($\omega$) is defined as follows [**125**]:

$$\alpha^2 F(\omega) = \frac{1}{2\pi N(E_F)} \sum_{qv} \frac{\gamma_{qv}}{\hbar\omega_{qv}} \delta(\omega - \omega_{qv}) \quad (17)$$

Here, $\omega_{qv}$ is the phonon frequency, while $\gamma_{qv}$ corresponds to the phonon linewidth for the phonon with wave vector $q$ and branch $v$.

$$\gamma_{qv} = 2\pi\omega_{qv} \sum_{i,j} \int_{BZ} \frac{d^3k}{\omega_{BZ}} |g_{qv}(k,i,j)|^2 \delta(\varepsilon_{k,i} - \varepsilon_F)\delta(\varepsilon_{k+q,j} - \varepsilon_F) \quad (18)$$

Here, $g_{qv}$ ($k$, $i$, $j$) and $\varepsilon_{q,i}$ represents the matrix of the EPC and the electronic energy, respectively.

The total EPC constants are calculated to be $\lambda = 1.18$ for ZrRuAs and $\lambda = 1.30$ for HfRuP, indicating that both materials belong to the strong-coupling conventional superconducting regime. By substituting the calculated $\lambda$ and $\omega_{\log}$ into **Equation 14**, the superconducting transition temperature ($T_c$) was obtained. The estimated $T_c$ as well as superconducting parameters of $\lambda$ and $\omega_{\log}$ are summarized in **Table 4**. Although HfRuP exhibits slightly stronger electron-phonon coupling, ZrRuAs shows a higher superconducting transition temperature (9.75 K) due to its larger logarithmic phonon frequency.

# 4 Conclusions

In summary, a detailed first-principles study has been carried out to investigate the structural, electronic, mechanical, bonding, optical, thermodynamic, topological, and superconducting properties of ZrRuAs and HfRuP. The optimized structural parameters are in good agreement with the available experimental values that confirms the reliability of the computational approach. The electronic band structures show that both compounds possess semimetallic characteristics. The inclusion of spin-orbit coupling modifies the band dispersion by lifting band degeneracies. The Fermi surface analysis further reveals complex electron and hole-like sheets which indicates anisotropic electronic behavior at low energy.

The calculated elastic constants satisfy the Born mechanical stability criteria which confirms that both compounds are mechanically stable. The values of Pugh's ratio, Poisson's ratio, and Cauchy pressure indicate ductile behavior while the low hardness values suggest that both materials are mechanically soft. The bonding analysis reveals mixed metallic, ionic, and covalent bonding characteristics which play an important role in stabilizing the crystal structures. Optical calculations show strong reflectivity and significant absorption in the ultraviolet region that suggests possible relevance for optoelectronic and reflective coating applications. The optical anisotropy is low for both the compounds. The thermodynamic parameters calculated under different temperatures and pressures confirm the thermodynamic stability of both compounds over the investigated range.

The absence of negative phonon modes confirms the dynamical stability of ZrRuAs and HfRuP. Electron-phonon coupling calculations show that the low-frequency phonon modes are mainly dominated by out-of-plane Ru vibrations and make the major contribution to superconductivity. The calculated electron-phonon coupling constants indicate that both compounds belong to the strong-coupling conventional superconducting regime. The estimated superconducting transition temperatures are 9.75 K for ZrRuAs and 9.03 K for HfRuP. The soft nature of the compounds suggests that moderate pressure can lead to substantial change in the electronic and lattice dynamic parameters. These, in turn, can change the superconducting transition temperature significantly [126-128]. Overall, the present study provides a comprehensive understanding of the unexplored physical properties of ZrRuAs and HfRuP and may help to support future experimental and theoretical investigations of topological superconducting materials.

## Acknowledgements

S.H.N. and R.S.I. acknowledge the research grant (1151/5/52/RU/Science-07/19-20) from the Faculty of Science, University of Rajshahi, Bangladesh, which partly supported this work. This work is dedicated to the cherished memory of the martyrs of the July-August 2024 movement in Bangladesh, whose sacrifices will forever inspire us.

## Data availability

The data sets generated and/or analyzed in this study are available from the corresponding author on reasonable request.

## Declaration of interest

The authors declare no competing interests.

## CRediT authorship contribution statement

**Jubair Hossan Abir:** Software, Methodology, Formal analysis, Data curation, Visualization, Writing-original draft, review & editing; **Tanvir Khan & Tauhidur Rahman:** Formal analysis, Visualization, Writing- draft, review & editing; **Md. Kamrul Hassan:** Software, Investigation, Data Curation; **Sraboni Saha Moly & Mst. Maskura Khatun:** Writing- draft, review; **Raihana Shams Islam:** Conceptualization, Supervision, Validation, Administration, Writing- Reviewing and Editing; **Saleh Hasan Naqib:** Conceptualization, Supervision, Validation, Administration, Writing- Reviewing and Editing.